\documentclass[13pt,twocolumn]{revtex4}
\usepackage{graphicx}    
\usepackage{graphicx}
\usepackage{amssymb}
\usepackage{verbatim}   
\usepackage{amsmath}    
\usepackage{color}      

\begin{document}
\title{``Blue energy'' from ion adsorption and electrode charging in sea- and river water}
\author{Niels Boon and Ren\'e van Roij} \affiliation{Institute for Theoretical Physics, Utrecht University, Leuvenlaan 4, 3584 CE Utrecht, The Netherlands}

\begin{abstract}
A huge amount of entropy is produced at places where fresh water and seawater mix, for example at river mouths. This mixing process is a
potentially enormous source of sustainable energy, provided it is harnessed properly, for instance by a cyclic charging and discharging process of
porous electrodes immersed in salt and fresh water, respectively [D. Brogioli, Phys. Rev. Lett. {\bf 103}, 058501 (2009)]. Here we employ a
modified Poisson-Boltzmann free-energy density functional to calculate the ionic adsorption and desorption onto and from the charged electrodes,
from which the electric work of a cycle is deduced. We propose optimal (most efficient) cycles for two given salt baths involving two canonical
and two grand-canonical (dis)charging paths, in analogy to the well-known Carnot cycle for heat-to-work conversion from two heat baths involving two isothermal
and two adiabatic paths. We also suggest a slightly modified cycle which can be applied in cases that the stream of fresh water is limited.
\end{abstract}

\maketitle
\onecolumngrid
{\em Just because there's a lot of it, doesn't mean it's interesting.}\\
Bob Evans about water (1998).\\
~\newline
{\em Charges are the invention of the devil.}\\
Bob Evans about ionic criticality (1998).
~\newline
\twocolumngrid
\section{Introduction} Where river water meets the sea, an enormous amount of energy is dissipated as a result of the irreversible mixing of fresh
and salt water. The dissipated energy is about 2 kJ per liter of river water, {\em i.e.} equivalent to a waterfall of 200m ~\cite{pattle}. It is
estimated that the combined power from all large estuaries in the world could take care of approximately 20\% of today's worldwide energy demand
\cite{wick}. Extracting or storing this energy is therefore a potentially serious option that our fossil-fuel burning society may have to embrace
in order to become sustainable. However, interesting scientific and technical challenges are to be faced. So far pressure-retarded osmosis (PRO)
\cite{levenspiel, loeb1975, loeb1976, thorsen, post} and reverse electrodialysis (RED) \cite{weinstein, veerman, post, post2008, dlugolecki} have been the two main and best-investigated
techniques in this field of so-called ``blue energy'', or salinity-gradient energy. In PRO the osmotic pressure difference
across a semi-permeable membrane is used to create a pressurised solution from incoming fresh and salt water, which is able to drive a turbine
\cite{levenspiel, loeb1975, loeb1976, thorsen, post}. In RED stacks of alternating cation- and anion-exchange membranes are used to generate an
electric potential difference out of a salinity gradient \cite{weinstein, veerman, post, post2008, dlugolecki}. These techniques enable the generation of (electrical) work at the expense of the mixing of streams with different salinity. Actually, PRO and RED can be thought of as the inverse processes of reverse osmosis and electrodialyses, where one has to supply (electrical) work in order to separate an incoming salt-water stream in a saltier and a fresher stream.\\

The applicability of  PRO and RED are currently being explored: a 1-2 kW
prototype plant based on PRO was started up in 2009 in Norway \cite{statkraft}, and a 5 kW RED device is
planned to be upscaled to a 50 kW demonstration project in The Netherlands \cite{redstack}. Interestingly, the bottleneck to large-scale applications
of both these techniques is often {\em not} the available fuel ---there is a lot of fresh and salt water--- but rather the very large membranes that
are required to operate at commercially interesting power outputs. Tailoring such membranes with a very high transport capacity and minimal
efficiency losses due to biofouling requires advanced membrane technology.

Recently, however, a solid-state device {\em without} membranes was constructed by Brogioli \cite{brogioli}, who directly extracts energy from salinity differences using porous carbon electrodes immersed in an aqueous electrolyte. Due to the huge internal surface of
porous carbon, of the order of $10^3$ m$^2$ per gram of carbon, the capacitance of a pair of electrolyte-immersed porous carbon electrodes can be
very large, allowing for large amounts of ionic charge to be stored in the diffuse part of the double layers of the electrolytic medium inside the pores
\cite{simon}. In fact, although the energy that is stored in the charged state of such large-area electrodes is somewhat lower than that in modern
chargeable batteries, the power uptake and power delivery of these ultracapacitors is comparable or even larger \cite{simon}. The capacitance of
these devices not only scales with the contact area between the electrode and the electrolyte, but also with the inverse distance between the electronic
charge on the electrode and the ionic charge in the diffuse part of the double layer, i.e. the capacitance increases with the inverse of the thickness of the ionic double layer. As a
consequence, the capacitance increases with increasing salinity, or, in other words, the potential increases at fixed electrode charge upon
changing the medium from salt to fresh water. This variability of the capacity was used by Brogioli \cite{brogioli}, and also more recently by Brogioli \emph{et al.}\cite{zhao}, to extract electric work from salinity gradients without membranes.
Although Sales \emph{et al.} showed that the combination of membranes and porous electrodes has some desirable advantages \cite{sales}, we will focus here on Brogioli's experiment.

The key concept of Ref.\cite{brogioli} is a four-stage cycle ABCDA of a pair of porous electrodes, together forming a capacitor, such that
\begin{enumerate} \item[(AB)] the two electrodes, immersed in sea water, are charged up from an initial state A with
low initial charges $\pm Q_A$ to a state B with higher charges $\pm Q_B$;
\item[(BC)] the salt water environment of the two electrodes is replaced by fresh water at fixed electrode charges $\pm Q_B$,
thereby increasing the electrostatic potential of the electrodes from $\pm\psi_B$ to $\pm\psi_C$;
\item[(CD)] the two highly charged electrodes, now immersed in fresh water in state C, are discharged back to $\pm Q_A$ in state D, and finally
\item[(DA)] the fresh water environment of the electrodes is replaced by salt water again, at fixed electrode charges $\pm Q_A$, thereby lowering the
electrode potentials to their initial values $\pm\psi_A$ in state A.
\end{enumerate}
This cycle, during which a net transport of ions from salt to fresh water takes place, renders the salt water fresher and the fresh water saltier
---although only infinitessimally so if the reservoir volumes are infinitely large. As a consequence, the ionic entropy has increased after a cycle has been completed, and the associated free-energy reduction of the combined
device and the two electrolyte reservoirs equals the electric work done by the device during the cycle, as we will see in more detail below.
Brogioli extrapolates an energy production of 1.6 kJ per liter of fresh water in his device \cite{brogioli}, equivalent to a waterfall of 160 m,
quite comparable to current membrane-based techniques. These figures are promising in the light of possible future large-scale blue-energy extraction. Together with the large volume of fresh and salt water at the river mouths of this planet, they also put an interesting and blessing
twist to Bob Evans' quotes at the beginning of this article.

Below we investigate the (free) energy and the performed work of electrolyte-immersed supercapacitors within a simple density
functional that gives rise to a modified Poisson-Boltzmann (PB) equation for the ionic double layers. By seeking analogies with the classic Carnot
cycle for heat engines with their maximum efficiency to convert heat into mechanical work given the two temperatures of the heat baths, we consider
modifications of Brogioli's cycle that may maximise the conversion efficiency of ionic entropy into electric work given the two reservoir salt
concentrations. Our modification does {\em not} involve the trajectories AB and CD of the cycle where the (dis)charging electrodes are in
diffusive contact with an electrolytic reservoir
---with the inhomogeneously distributed salt ions ``properly'' treated grand-canonically as often advocated by Bob Evans \cite{evans1979,tarazona1987,evans1989}.
In fact, we will argue that the grand-canonical trajectories AB and CD at constant ionic chemical potential are the analogue of the isotherms in
the Carnot cycle. Rather we consider to modify the constant-charge trajectories BC and DA (which correspond to isochores in a heat-engine as we
will argue) by a continued (dis)charging process of the electrodes at a constant number of ions (which corresponds to an adiabatic (de)compression
in the heat engine). In other words, we propose to disconnect the immersed electrodes from the ion reservoirs in BC and DA, treating the salt ions
canonically while (dis)charging the electrodes, thereby affecting the ion adsorption and hence the bulk concentration from salty to fresh (BC) and
{\em vice versa} (DA). Finally, we will consider a (dis)charging cycle in the (realistic) case of a finite volume of available fresh water, such
that the ion exchange process renders this water brackish; the heat-engine analogue is a temperature rise of the cold bath due to the uptake of
heat.\\

Similar cycles were already studied theoretically by Biesheuvel \cite{biesheuvel2009}, although not in this context of osmotic power but its reverse, capacitive desalination. The ``switching step'' in Biesheuvel's cycle, where the system switches from an electrolyte with a low salt concentration to an electrolyte with a higher salt concentration, appears to be somewhat different from our proposal here, e.g. without a direct heat-engine analogue.\\

\section{System and Thermodynamics} We consider two electrodes, one carrying a charge $Q$ and the other a charge $-Q$. The electrodes, which can
charge and discharge by applying an external electric force that transports electrons from one to the other, are both immersed in an aqueous
monovalent electrolyte of volume $2V$ at temperature $T$. We denote the number of cations and anions in the volume $2V$ by $2N_+$ and $2N_-$,
respectively. Global charge neutrality of the two electrodes and the electrolyte in the volume $2V$ is guaranteed if $2N_+=2N_-$. If the two
electrodes are separated by a distance much larger than the Debye screening length ---a condition that is easily met in the experiments of
Ref.\cite{brogioli}--- then each electrode and its surrounding electrolyte will be separately electrically neutral such that $Q/e=N_--N_+$, where
$e$ is the proton charge and where we assume $Q>0$ without loss of generality. Note that this ``local neutrality'' can only be achieved provided
$Q/e\leq N_++N_-\equiv N$, where the extreme case $Q/e=N$ corresponds to an electrode charge that is so high that all $2N_-$ anions in the volume
$2V$ are needed to screen the positive electrode and all $2N_+$ cations to screen the negative one. For $Q/e\leq N$, which we assume from now on,
we can use $Q$ and $N$ as independent variables of a neutral system of the positive electrode immersed in an electrolyte of volume $V$ at
temperature $T$, the Helmholtz free energy of which is denoted by $F(Q,N,T,V)$. At fixed volume and temperature we can write the differential of
the free energy of the positive electrode and its electrolyte environment as
\begin{equation}
\mathrm{d}F = \mu\mathrm{d}N + \Psi\mathrm{d}Q, \label{dF}
\end{equation}
with $\mu=(\mu_++\mu-)/2$ the average of the ionic chemical potentials $\mu_{\pm}$ and $\Psi$ the electrostatic potential of the electrode. The
last term of Eq.(\ref{dF}) is the electric work {\em done on} the system if the electrode charge is increased by $dQ$ at fixed $N$, and hence the
electrostatic work {\em done by} the electrode system is ${\rm d}W\equiv-\Psi {\rm d}Q$. Given that $F$ is a state function, such that $\oint
dF=0$ for any cycle, the total work {\em done by} the system during a (reversible) cycle equals
\begin{equation}\label{W}
W\equiv\oint \mathrm{d}W=-\oint \Psi \mathrm{d}Q=\oint \mu \mathrm{d}N.
\end{equation}
In order to be able to {\em calculate} $W$ we thus need explicit cycles {\em and} the explicit equations-of-state $\mu(Q,N,T,V)$ and/or
$\Psi(Q,N,T,V)$, for which we will use a simple density functional theory to be discussed below.

\begin{figure}[]
\centering
\includegraphics[width = 7cm]{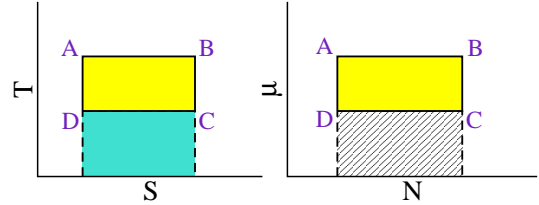}
\caption{A schematic comparison of (a) the entropy-temperature $(S,T)$ representation of the Carnot cycle of a gas that extracts mechanical work
from exchanging heat (and thus entropy) between heat baths at two fixed temperatures and (b) the $(N,\mu)$ representation (see text) of a cycle to
extract electric work from exchanging ions between ionic reservoir at two fixed chemical potentials.}
\label{fig:carnotcompare}
\end{figure}

However, before performing these explicit calculations a general statement can be made, because there is an interesting analogy to be made with
mechanical work $W_m=\oint pdV$ {\em done by} a fixed amount of gas at pressure $p$ that cyclically changes its volume and entropy (by exchanging
heat). In that case the differential of the thermodynamic potential reads $dU=TdS - pdV$ with $U$ a state function denoting the internal energy.
Since $\oint dU=0$ we then find $W_m=\oint TdS$. If the exchange of heat takes place between two heat baths at given high and low temperatures
$T_H$ and $T_L$,  it is well known that the most-efficient cycle ---the cycle that produces the maximum work per adsorbed amount of heat from the
hotter bath--- is the Carnot cycle with its two isothermal and two adiabatic (de-)compressions \cite{kittel}. If we transpose all the variables
from the gas performing mechanical work to the immersed electrodes performing electric work, we find $U\Leftrightarrow F$, $S\Leftrightarrow N$, $T\Leftrightarrow\mu$,
$V\Leftrightarrow Q$, and $-p\Leftrightarrow\Psi$, where all pairs preserve the symmetry of being both extensive or both intensive. The analogue
of high and low temperatures are thus high and low ionic chemical potentials $\mu_H$ and $\mu_L$ (corresponding to sea and river water,
respectively), the analogue of the isothermal volume change is thus the (dis)charging at constant $\mu$, and the analogue of an adiabatic volume
change is (dis)charging at constant $N$. Therefore, the analogue of the most efficient gas cycle is the electric cycle consisting of
(grand)canonical (dis)charging processes. Indeed, the trajectories (AB) and (CD) of the experimental cycle of Ref.\cite{brogioli}, as discussed in
section I, are of a grand-canonical nature with the electrode in contact with a salt reservoir during the (dis)charging. However, the processes
(BC) and (DA) take place at constant $Q$, i.e. they are equivalent to isochores in a gas cycle, instead of adiabats. Efficiency is thus to be
gained, at least in principle, by changing BC and DA into canonical charging processes. Whether this is experimentally easily implementable is, at
this stage for us, an open question that we will not answer here.

For the most efficient cycles, which are schematically shown in  Fig.\ref{fig:carnotcompare} in the $(S,T)$ and the $(N,\mu)$ representation, we
can easily calculate the work performed during a cycle. For the mechanical work of the gas one finds $W_m=\Delta S\Delta T$, with $\Delta
T=T_H-T_L$ the temperature difference and $\Delta S$ the entropy that is exchanged between the heat baths during the isothermal compression and
decompression. The analogue for the work $W$ delivered by the electrode is given by $W=\Delta\mu\Delta N$, with $\Delta\mu=\mu_H-\mu_L$ and
$\Delta N$ the number of exchanged ions between the reservoirs during the grand-canonical (dis)charging processes. This result also follows
directly from Eq.(\ref{W}). Below we will calculate $\Delta N$ and hence $W$ from a microscopic theory. Moreover, we will also consider several other types of cycles.

\begin{figure}[]
\centering
\includegraphics[width = 7cm]{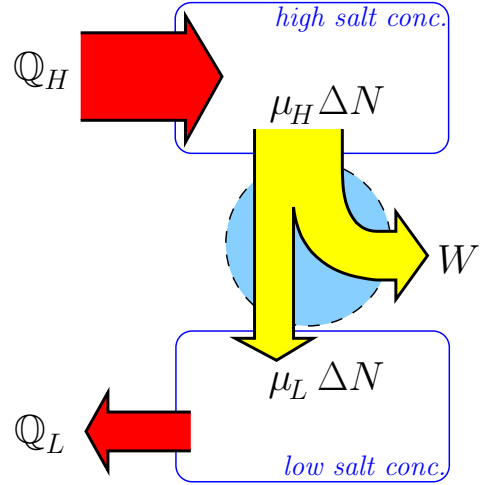}
\caption{A scheme which illustrates the generation of reversible work due to a flow of $\Delta N$ particles through the device from a reservoir with a high salt concentration to a reservoir with a low salt concentration. The difference in chemical potential between the two reservoirs enables for the extraction of an amount of reversible work $W=(\mu_H-\mu_L) \Delta N$, which equals the decrease of the net free energy of the reservoirs. The increase of entropy due to this controlled mixing process is caused by an inflow $\mathbb{Q} = \mathbb{Q_H} - \mathbb{Q_L}$ of heat from the atmosphere, which means that the system tends to cool down as it performs work. Note that the division of this heat into two flows of different directions is purely suggestive, as in general there is a transport of heat trough the device as well.}
\label{fig:flows}
\end{figure}

In the context of the thermodynamics that we discuss here, it is also of interest to analyse the ``global'' energy flow that gives rise to the work $W$  that the immersed porous electrodes deliver per (reversible) cycle. For this analysis it is crucial to realise that the device and the two salt reservoirs at chemical potentials $\mu_H$ and $\mu_L$ are considered to be at constant temperature $T$ throughout, which implies that they are thermally coupled to a heat bath (that we call the ``atmosphere'' here for convenience) at temperature $T$. We will show that with every completed cycle, during which $\Delta N>0$ ions are being transported from the sea to the river water, a net amount of heat $\mathbb{Q}>0$ flows from the atmosphere to the two salt reservoirs, and that $W=\mathbb{Q}$ in the limit that the ion clouds do not store potential energy due to multi-particle interactions. This may at first sight contradict Kelvin's statement of the Second Law (``no process is possible whose sole result is the complete conversion of heat into work'' \cite{blundell}), but one should realise that the cycle {\em also} involves the transport of ions from the sea to the river; the word ``sole'' in Kelvin's statement is thus crucial, of course. The analysis is based on the entropy changes $\Delta S_d$, $\Delta S_H$ and $\Delta S_L$ of the device, the highly-concentrated salt reservoir and the one with low salt concentration, respectively, upon the completion of a cycle.  Given that the device returns to its initial state after a complete cycle, its entropy change vanishes and $\Delta S_d=0$. This implies that the device, at its fixed temperature, does not adsorb or desorb any net amount of heat. During a cycle the ``river'' gains $\Delta N$ ions, and hence its (Helmholtz or Gibbs) free energy changes by $\Delta F_L=\mu_L \Delta N$, while the ``sea'' loses $\Delta N$ ions such that $\Delta F_H=-\mu_H\Delta N$. Now the basic identity $F = E - TS$ implies that $\Delta F_H =  - \epsilon \Delta N - T \Delta S_H$ and $\Delta F_L = \epsilon \Delta N - T \Delta S_L$, where $\epsilon = E/N$ is the average energy (or enthalpy if $F$ denotes the Gibbs free energy) per particle. We assume $\epsilon$ to be independent of density, which physically corresponds to the case that there are no multi-particle contributions to the internal energy of the reservoirs, as is the case for hard-core systems or ions treated within Poisson-Boltzmann theory as ideal gases in a self-consistent field. The total energy in the reservoirs therefore remains constant during mixing, such that the entropy changes of the salt reservoirs are $T \Delta S_H= (\mu_H - \epsilon) \Delta N$ and $T \Delta S_L= -(\mu_L - \epsilon) \Delta N$. As a consequence of the global preservation of entropy in the reversible cycle, the ion exchange actually drives a heat exchange whereby the sea extracts a net amount of heat $\mathbb{Q}_H=T\Delta S_H$ from the atmosphere, while the river dumps a net amount of heat $\mathbb{Q}_L=-T\Delta S_L$ into the atmosphere. Of course the transport of ions itself is also accompanied with a heat exchange in between the reservoirs, the only relevant flow is therefore the net flow of heat out of the atmosphere, which is $\mathbb{Q}=\mathbb{Q}_H-\mathbb{Q}_L=\Delta\mu\Delta N=W$.
The energy flow and the particle flow of the device and reservoirs are tentatively illustrated in Fig.~\ref{fig:flows}, where one should realise that the distribution of the heat flow from the atmosphere into the sea ($\mathbb{Q}_H$) and the river ($-\mathbb{Q}_L$) depends on the heat-flow from river to sea or {\em vice versa}, which we have not considered here in any detail; {\em only} the net heat flow $\mathbb{Q}_H-\mathbb{Q}_L$ is fixed by global thermodynamic arguments. This identification of $\mathbb{Q}$ with $W$ would have the interesting implication that the conversion of this work into heat again, e.g. by using it to power a laptop, would {\em not} contribute to (direct) global warming since the released heat has previously been taken out of the atmosphere\cite{biesheuvellaptop}. It is not clear to us, however, to what extent this scenario is truly realistic and relevant, given that rivers, seas, and the atmosphere are generally {\em not} in thermal equilibrium such that other heat flows are to be considered. In this study we do not consider the heat fluxes at all, and just consider systems that are small enough for the temperature to be fixed.

\section{Microscopic model and density functional theory} 
\begin{figure}[]
\centering
\includegraphics[width = 8cm]{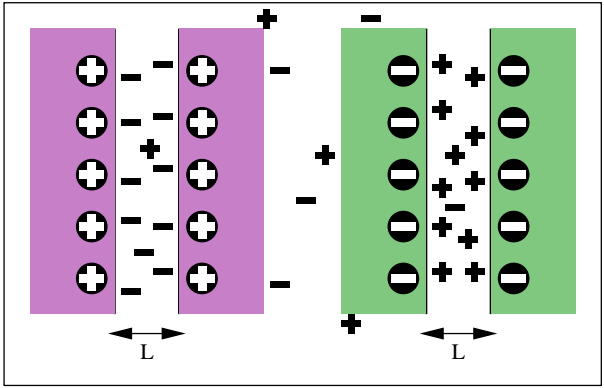}
\caption{A sketch of the two electrodes under consideration, one positively charged and the other one negatively charged, both in contact with an
electrolyte with a compensating ionic charge. The porosity of the electrodes is modeled by a slit of width $L$ filled with electrolyte in between
two solid half spaces with surface charge density $e\sigma$ for the positive electrode and $-e\sigma$ for the negative one, as represented by the
encircled plus and minus signs. The ions in the gap, represented by bare plus and minus signs, are free to migrate throughout the pores and also
to the volume outside the electrodes towards the other electrode.} \label{fig:electrodes}
\end{figure}

In order to calculate $\mu(Q,N,T,V)$ and $\Psi(Q,N,T,V)$ of a charged
electrode immersed in an electrolyte of volume $V$, we need a microscopic model of the electrode and the electrolyte. We consider a positively
charged porous electrode with a total pore volume $V_e$, total surface area $A$, and typical pore size $L$. We write the total charge of the positive
electrode as $Q=e\sigma A$ with $\sigma$ the number of elementary charges per unit area. The negative electrode is the mirror image with an
overall minus sign for charge and potential, see also Fig.\ref{fig:electrodes}. The volume of the electrolyte surrounding this electrode is
$V=V_e+V_o$, with $V_o$ the volume of the electrolyte outside the electrode. The electrolyte consists of (i) water, viewed as a dielectric fluid
with dielectric constant $\epsilon$ at temperature $T$ , (ii) an (average) number $N_-=(N+Q/e)/2$ of anions with a charge $-e$ and (iii) an (average) number $N_+=(N-Q/e)/2$ of cations with a charge $+e$. The finite pore size $L$ inside the electrodes is taken into account here only qualitatively by regarding a geometry of two laterally
unbounded parallel half-spaces representing the solid electrode, both with surface charge density $e\sigma$, separated by a gap of thickness $L$
filled with the dielectric solvent and an inhomogeneous electrolyte characterised by concentration profile $\rho_{\pm}(z)$. Here $z$ is the
Cartesian coordinate such that the charged planes are at $z=0$ and $z=L$. The water density profile $\rho_w(z)$ is then, within a simple
incompressibility approximation $(\rho_w(z)+\rho_+(z)+\rho_-(z))v=1$ with $v$ a molecular volume that is equal for water and the ions, given by
$\rho_w(z)=1/v-\rho_+(z)-\rho_-(z)$. If the electrolyte in the gap is in diffusive contact with a bulk electrolyte with chemical potentials
$\mu_+$ and $\mu_-$ of the cations and anions, we can write the variational grand-potential as a functional $\Omega[\rho_+,\rho_-]$ given by
\begin{eqnarray}
\frac{\Omega[\rho_\pm]}{Ak_BT}&=&\int_0^{L/2} \mathrm{d}z \left[\rho_+(z)\Big(-1+\ln \rho_+(z) \Lambda_+^3 -
\frac{\mu_+}{k_BT}\Big)\right.\nonumber\\
&&\,\,\,\,\,+\rho_-(z)\Big(-1+\ln \rho_-(z)\Lambda_-^3 -
\frac{\mu_-}{k_BT}\Big)\nonumber\\
&&\,\,\,\,\, + \rho_w(z)\Big(-1+\ln\rho_w(z)v\Big) \nonumber\\
&&\,\,\,\,\,+\left.\frac{\phi(z)q(z)}{2}\right].\label{func}
\end{eqnarray}
Here the first two lines denote the ideal-gas grand potential of the two ionic species,
with $\Lambda_\pm$ the ionic thermal wavelengths. The third
line is the ideal water-entropy, which effectively accounts for ionic excluded volume interactions as it restricts the total local ion
concentration to a maximum equal to $1/v$ ---of course we could have taken the much more accurate hard-sphere functionals closer to Bob Evans'
heart to account for steric repulsions \cite{tarazona1985,rosenfeld1989,evans1991,roth2002, dicaprio}, but for now we are satisfied with the more qualitative lattice-gas-like description of packing
\cite{bikerman, bazant2009,borukhov}. The last line of Eq.(\ref{func}) denotes the mean-field approximation of the electrostatic energy in terms of the total charge
number density $q(z)=\rho_+(z)-\rho_-(z)+\sigma(\delta(z)+\delta(z-L))$ and the electrostatic potential $\psi(z)=k_BT\phi(z)/e$. Note that
$\psi(z)$ is a functional of $\rho_{\pm}(z)$ through the Poisson equation $\phi''(z)=-4\pi\lambda_Bq(z)$ with $\lambda_B=e^2/\epsilon k_BT$ the
Bjerrum length of water, and that $\psi(0)\equiv\Psi$ is the electrode potential. A prime denotes a derivative with respect to $z$.

The Euler-Lagrange equations $\delta\Omega/\delta\rho_{\pm}(z)=0$ that describe the equilibrium concentration profiles yield $\mu_\pm\equiv k_BT\ln\left(\rho_s\Lambda_{\pm}^3 / (1-\eta_0)\right)$, with $\rho_s$ the bulk reservoir salt concentration and $\eta_0=2\rho_sv$ the ionic packing fraction in the reservoir, where $\phi(z)=0$. When the Euler-Lagrange equations are combined with the Poisson equation, the modified Poisson-Boltzmann (PB) equation with boundary conditions (BCs)
\begin{eqnarray}
\phi''(z) &=& \frac{\kappa^2 \sinh\phi(z)}{1-\eta_0 + \eta_0\cosh \phi(z)}; \label{modpb}\\
\left.\phi'(z)\frac{}{}\right|_{z=0} &=& -4\pi\lambda_B\sigma;\label{bc1}\\
\left.\phi'(z)\frac{}{}\right|_{z=\frac{L}{2}} &=& 0,\label{bc2}
\end{eqnarray}
with $\kappa^{-1}=(8\pi\lambda_B\rho_s)^{-1/2}$ the Debye screening length, is found. BC (\ref{bc1}) follows from Gauss' law on the surface of the electrode, and BC (\ref{bc2}) from charge-neutrality and the symmetry with respect to the midplane of the gap. Note that this equation with accompanying BC's was already studied in Ref.\cite{kraljiglic,paunov, paunov1996}. Eq.(\ref{modpb}) reduces to the standard PB equation if $v=0$, and the large-gap case $L\rightarrow\infty$ was studied in Ref.\cite{borukhov}. Eq. (\ref{modpb}) with its BC's (\ref{bc1}) and (\ref{bc2}) forms a closed set, and once its solution is found, numerically in general or analytically in the special case that $\eta_0=0$ and $\kappa L\rightarrow\infty$, the required equation of state of the electrode potential follows from $\Psi(\rho_s,\sigma)=k_BT\phi(0)/e$. Moreover, the equilibrium density profiles can be used to calculate the cationic and anionic adsorption, i.e. the excess number of ions per unit surface area, defined by
\begin{equation}
\Gamma_\pm(\sigma,\rho_s) = \int_{z=0}^{z=L/2}~\mathrm{d}z~ \Big(\rho_\pm(z) -  \rho_s\Big).\label{gampm}
\end{equation}
Note that we integrate the profile up to $z=L/2$ as required, and that our ``local charge neutrality'' assumption implies that $\sigma =
\Gamma_-(\sigma,\rho_s) - \Gamma_+(\sigma,\rho_s)$. Interestingly, the total surface excess of ions, defined by
 \begin{equation}
\Gamma(\sigma,\rho_s) = \Gamma_+(\sigma,\rho_s) + \Gamma_-(\sigma,\rho_s),\label{gam}
\end{equation}
is related to the total number of ions in the volume $V=V_e + V_o$ by
\begin{equation}
N=2\rho_s V + A\Gamma(\rho_s,\sigma).\label{nrions}
\end{equation}
Below we will use expression (\ref{gampm}) and (\ref{gam}) for $\Gamma(\rho_s,\sigma)$ to calculate $N(\rho_s,\sigma)$ from Eq.(\ref{nrions}), {\em or} to calculate
$\rho_s(N,\sigma)$ by solving Eq.(\ref{nrions}) for $\rho_s$ at given $N$ and $\sigma$.

Before discussing our numerical results, it is useful to consider the limiting case $\kappa L\gg 1$ and $\eta_0\ll1$, which is in fact the classic
Gouy-Chapman (GC) problem of a single, planar, charged wall in contact with a bulk electrolyte of point ions. In this case the PB equation can be
solved analytically \cite{israelachvili,vanroij2010}, and the resulting total adsorption is given by
\begin{eqnarray}
\Gamma_{GC}(\sigma,\rho_s) &=& \sqrt{\sigma^2 +\frac{\kappa^2}{4 \pi^2\lambda_B^2}} - \frac{\kappa}{2 \pi\lambda_B}\nonumber\\
&=&\left\{
\begin{array}{ll}\displaystyle \frac{\sigma^2}{2\sigma^*}, &\,\,\,\sigma\ll\sigma^{*};\\ \sigma, &\,\,\, \sigma\gg\sigma^{*},\end{array}\right.\label{gc}
\end{eqnarray}
with the crossover surface charge $\sigma^*=\frac{\kappa}{2\pi\lambda_B}$. The crossover behavior from $\Gamma\propto\sigma^2$ at low $\sigma$ to
$\Gamma=\sigma$ at high $\sigma$ signifies a qualitative change from the linear screening regime, where the double layers exchange co- for
counter ions keeping the total ion concentration fixed (such that $\Gamma$ is small), to the nonlinear screening regime where counterion
condensation takes place. For the alleged most-efficient (dis)charging cycle of current interest, operating between two ionic reservoirs with a high salt
concentration $\rho_s=\rho_H$ and a low one $\rho_s=\rho_L$ such that $\Delta\mu=\mu_H-\mu_L=k_BT\ln(\rho_H/\rho_L)$, and for which we argued
already that the electric work per cycle reads $W=\Delta N\Delta\mu$, the GC result (\ref{gc}) allows for the calculation of the ionic uptake
$\Delta N=A(\Gamma(\rho_H,\sigma_B)-\Gamma(\rho_H,\sigma_A))$ during the grand-canonical charging from a low charge density $\sigma_A$ to a high
one $\sigma_B$. In the limit of highly charged surfaces we thus find $\Delta N=A(\sigma_B-\sigma_A)$, and hence the optimal work per unit area
within the GC limit reads
\begin{eqnarray}
\frac{W_{GC}}{A} &=&  k_BT(\sigma_B - \sigma_A) \ln \frac{\rho_H}{\rho_L}\hspace{0.3cm} \mbox{for\ }\sigma_A\gg\sigma^*. \nonumber\\
\label{work}
\end{eqnarray}
With the typical numbers $\sigma_B-\sigma_A$ of the order of nm$^{-2}$, $A=10^3$ m$^2$ per gram of porous carbon, and
$\rho_H/\rho_L=100$ one arrives at $W_{GC}=10$ J per gram of carbon. Interestingly, this is substantially higher than Brogioli's experimental findings of only $\approx 20$ mJ/gram per cycle. We will discuss this difference below. The GC limit also yields an
analytic expression for the surface potential $\Psi_{GC}$ given by
\begin{equation}
\frac{e\Psi_{\rm{GC}}}{k_BT} = 2\rm{arcsinh}\Big(\frac{\sigma}{\sigma^*}\Big)\simeq
\left\{\begin{array}{ll}\displaystyle\frac{\sigma}{2\sigma^*},&\,\,\,\sigma\ll\sigma^{*};\\
\displaystyle 2\ln\frac{2\sigma}{\sigma^*},&\,\,\,\sigma\gg\sigma^{*}.\end{array}\right. \label{psigc}
\end{equation}
With typical Debye lengths $\kappa^{-1}\simeq 1$ nm and $\lambda_B=0.72$ nm we find for the typical crossover surface charge density
$\sigma^*\simeq 0.2$ nm$^{-2}$ which corresponds to a surface potential $\Psi\simeq 50$ mV. For completeness we also mention the
differential capacitance $C={\rm d}Q/{\rm d}\Psi$ of an immersed electrode, which within the GC limit was already written by Chapman \cite{chapman} as
\begin{equation}
\frac{C_{GC}}{A}=  \left(\frac{1}{e}\frac{\rm{d}\Psi_{GC}}{\rm{d}\sigma}\right)^{-1} = \frac{\kappa \epsilon}{4\pi} \cosh\left[\frac{e\Psi}{2 k_BT}\right],
\end{equation}
which corresponds to a two-plate capacitor with spacing $\kappa^{-1}/\cosh(e \Psi/ (2 k_BT))$ in
a dielectric medium characterised by its relative dielectric constant $\epsilon$. This result shows that the capacity indeed increases with the salt concentration, in agreement
with observations that the electrode potential rises at fixed charge upon sweetening the surrounding water \cite{brogioli}.

Useful insights can be obtained from these analytic GC expressions. Moreover, practical linear approximations \cite{ettelaie, biesheuvel2001} and even (almost) analytical solutions \cite{verweyoverbeek, ninham, torres} for the PB equation exist in the case of small pore size $L$. Nevertheless, the pointlike nature of the ions gives rise to surface concentrations of counterions that easily become unphysically large, e.g. far beyond 10M for the parameters of interest here. For this reason we consider the steric effects through the finite ionic volume $v$ and the finite pore size $L$ below, at the expense of some numerical
effort.

\section{Numerical results}
The starting point of the explicit calculations is the numerical solution of Eq.~(\ref{modpb}) with BC's~(\ref{bc1}) and~(\ref{bc2}) on a discrete
$z$-grid of $5000$ equidistant points on the interval $z\in[0,L/2]$, which we checked to be sufficient for all values of $\kappa L$ that we
considered. Throughout the remainder of this text we set $v=a^3$ with $a=0.55$ nm, which restricts the total local ion concentration $\rho_+(z)+\rho_-(z)$ to a physically
reasonable maximum of 10 M. The Bjerrum length of water is set to $\lambda_B=0.72$ nm.

\begin{figure}[]
\centering
\includegraphics[width = 7cm]{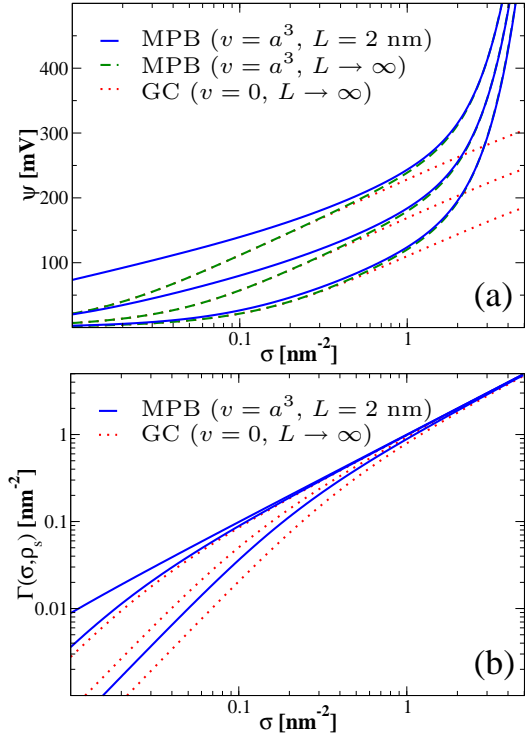}
\caption{(a) The surface potential $\Psi$ and (b) the ionic adsorption $\Gamma$, both as a function of the surface charge density $\sigma$, for a
porous electrode with a pore size $L=2$ nm immersed in an aqueous electrolyte of monovalent cations and anions at
reservoir salt concentrations $\rho_s=\{1\rm{mM},~10\rm{mM},~100\rm{mM}\}$ from top to bottom. The labels denote results stemming from the modified Poisson-Boltzmann (MPB) and Gouy-Chapman theory, with the corresponding molecular volume $v$ and the (finite) poresize $L$.}
\label{fig:potchargerel}
\end{figure}

We first consider a positive electrode immersed in a huge ($V\gg V_e$) ionic bath at a fixed salt concentration $\rho_s$, such that the ions can
be treated grand-canonically. In Fig.~\ref{fig:potchargerel} we plot (a) the electrode potential $\Psi$ and (b) the total ion adsorption $\Gamma$,
both as a function of the electrode charge number density $\sigma$, for three reservoir salt concentrations $\rho_s=$1, 10, and 100 mM from top to
bottom, where the full curves represent the full theory with pore size $L=2$ nm, the dashed curves the infinite pore limit $\kappa L\gg 1$, and
the dotted curve the analytic Gouy-Chapman expressions (for $\kappa L\gg 1$ and $v=0$) of Eqs.(\ref{gc}) and (\ref{psigc}). The first observation
in Fig.~\ref{fig:potchargerel}(a) is that GC theory breaks down at surface charge densities beyond $~1 e$ nm$^{-2}$, where steric effects prevent
too dense a packing of condensing counterions such that the actual surface potential rises much more strongly with $\sigma$ than the logarithmic
increase of GC theory (see Eq.(\ref{psigc})). This rise of the potential towards $\simeq 1$ V may induce (unwanted) electrolysis in experiments,
so charge densities exceeding, say, 5$e$ nm$^{-2}$ should perhaps be avoided. A second observation is that the finite pore size $L$ hardly affects
the $\Psi(\sigma)$ relation for $\sigma>1$ nm$^{-2}$, provided the steric effects are taken into account. The reason is that the effective
screening length is substantially smaller than $L$ in these cases due to the large adsorption of counterions in the vicinity of the electrode. A
third observation is that the full theory predicts,  for the lower salt concentrations $\rho_s=1$ and 10 mM, a substantially larger $\Psi$ at low
$\sigma$, the more so for lower $\rho_s$. This is due to the finite pores size, which is {\em not} much larger than $\kappa^{-1}$ in these cases,
such that the ionic double layers must be distorted: by increasing $\Psi$ a Donnan-like potential is generated in the pore that attracts enough
counterions to compensate for the electrode charge in the small available volume. Interestingly, steric effects do {\em not} play a large role for
$\Gamma(\sigma)$ in Fig.~\ref{fig:potchargerel}(b), as the full curves of the full theory with $v=a^3$ are indistinguishable from the full theory
with $v=0$. The finite pore size appears to be more important for $\Gamma(\sigma)$, at least at first sight, at low $\sigma$, where $\Gamma_{GC}$
appears substantially lower than $\Gamma$ from the full calculation in the finite pore. However, this is in the linear regime where the adsorption
is so small that only the logarithmic scale reveals any difference; in the nonlinear regime at high $\sigma$ all curves for $\Gamma$ coincide and
hence the GC theory is accurate to describe the adsorption.

\begin{figure}[]
\centering
\includegraphics[width = 8cm]{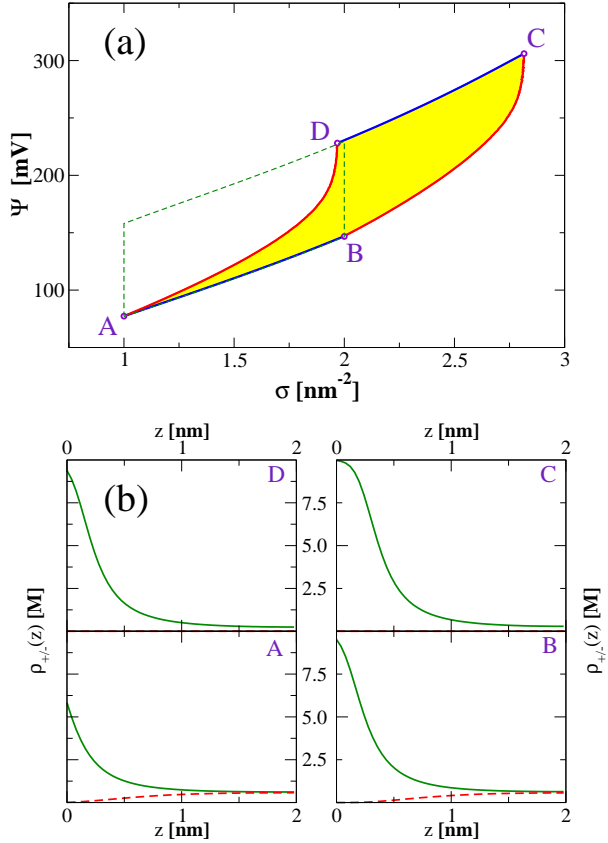}
\caption{(a) A cycle ABCDA of the surface potential $\Psi$ and the surface charge density $\sigma$ of an electrode with pore size $L=4$ nm,
charging up (AB) in contact with an electrolyte reservoir with a high salt concentration $\rho_s=\rho_H=0.6\rm{M}$ from $\sigma_A=1$ nm$^{-2}$ to
$\sigma_B=2$ nm$^{-2}$, discharging (CD) in contact with an electrolyte at low salt concentration $\rho_s=\rho_L=0.024\rm{M}$, while being
disconnected from the reservoirs (so with a fixed number of ions) during the additional charging (BC) and discharging (DA). The dashed cycle ABC'D'A, for which the surface charge remains fixed upon the transfer between the two reservoirs, resembles the cycle of the experiments of Ref.\cite{brogioli}. (b) Counter- and co-ion concentration
profiles $\rho_-(z)$ (full curves) and $\rho_+(z)$ (dashed curves), respectively, in the four states A, B, C, and D of (a).} \label{fig:cyclus}
\end{figure}

We now consider the (reversible) $\Psi$-$\sigma$ cycle ABCDA shown in Fig.\ref{fig:cyclus}(a), for an electrode with pore sizes $L=4$ nm that
operates between two salt reservoirs at high and low salt salt concentrations $\rho_s=\rho_H=0.6$ M (sea water) and $\rho_s=\rho_L=0.024$ M (river
water), respectively, such that $\Delta\mu/k_BT= 3.3$. For simplicity we set $V_o=0$ such that the total electrolyte volume
equals the pore volume $V_e=AL/2$. The trajectory AB represents the charging of the electrode from an initial charge density $\sigma_A=1$ nm$^{-2}$
to a final charge density $\sigma_B=2$ nm$^{-2}$ at $\rho_s=\rho_H$, which involves an increase in the number of ions per unit area $\Delta
N/A=\Gamma(\sigma_B,\rho_H)-\Gamma(\sigma_A,\rho_H)=0.7$ nm$^{-2}$ using Eqs.(\ref{gam}) and (\ref{nrions}) which we calculate numerically with
(\ref{gampm}). The trajectory BC is calculated using the fixed number of particles in state $B$, $N=N_B=2\rho_H V+A\Gamma(\rho_H,\sigma_B)$,
calculating a lower and lower value for $\rho_s$ for increasing $\sigma$'s using Eq.(\ref{nrions}) until $\rho_s=\rho_L$ at $\sigma=\sigma_C=2.81$ nm$^{-2}$.
Then the discharging curve CD, at fixed $\rho_s=\rho_L$ is traced from surface charges $\sigma_C$ down to $\sigma_D=1.97$ nm$^{-2}$ for which
$\Gamma(\sigma_D,\rho_L)-\Gamma(\sigma_C,\rho_L)=-\Delta N/A$, i.e. the discharging continues until the number of expelled ions equals their
uptake during the charging process AB. The final trajectory, DA, is characterised by the fixed number of particles in state D (which equals that
in A), and is calculated by numerically finding higher and higher $\rho_s$-values from Eq.(\ref{nrions}) for surface charges $\sigma$ decreasing
from $\sigma_D$ to $\sigma_A$, where $\rho_s=\rho_H$ at $\sigma=\sigma_A$ such that the loop is closed. Note that all four trajectories involve
numerical solutions of the modified Poisson-Boltzmann problem and some root-finding to find the state points of interest, and that the loop is
completely characterised by $\rho_H$, $\rho_L$, $\sigma_A$, and $\sigma_B$. Fig.\ref{fig:cyclus}(b) shows the concentration profiles of the anions
(full curve) and cations (dashed curves) in the states A, B, C, and D, (i) showing an almost undisturbed double layer in A and B that reaches local
charge neutrality and a reservoir concentration $\rho_{\pm}(z)=\rho_H$ in the center of the pore, (ii) an increase of counterions at the expense of
a decrease of coions in going from B to C by a trade off with the negative electrode, accompanied by the saturation of counterion concentration at
10 M close to the electrode in state C and the (almost) complete absence of co-ions in the low-salt states C and D, and (iii) the trading of
counterions for coions from D to A at fixed overall ion concentration.

The work done during the cycle ABCDA follows from either the third or the fourth term of Eq.(\ref{W}), yielding $W/A=2.3 k_BT$ nm$^{-2}$ or,
equivalently, $W/\Delta N=3.3 k_BT$ for the present set of parameters. The enclosed area of the cycle ABCDA in Fig.~\ref{fig:cyclus} corresponds to the amount of extracted work (up to a factor $(e A)$~), and equals the net decrease of free energy of the reservoirs.

\begin{figure}[]
\centering
\vspace*{1 cm}
\includegraphics*[width = 8cm]{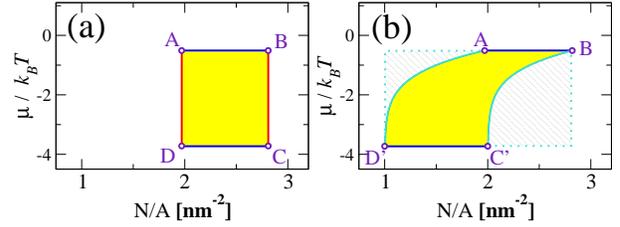}
\caption{The same two cycles as shown in Fig.\ref{fig:cyclus}(a), but now in the number of ions and chemical potential ($N$-$\mu$) representation
(see text), showing the constant-$N$ trajectories of BC and DA in (a) and a much larger spread in $N$ for the cycle ABC'D'A in (b). Whilst the solid areas represent the work performed during a cycle, the dashed areas in (b) denote losses
due to the non-optimal characteristics of the cycle for the given range of $N$. The enclosed areas give the amount of work that is extracted per cycle and per unit electrode area.} \label{fig:cyclus2}
\end{figure}

In order to compare the presently proposed type of cycle ABCDA with the type  used in the experiments of Brogioli \cite{brogioli}, where
``isochores'' at constant $\sigma$ rather than ``adiabats'' at constant $N$ were used to transit between the two salt baths, we also numerically study
the dashed cycle ABC'D'A of Fig.\ref{fig:cyclus}(a). This cycle has exactly the same trajectory AB characterised by $(\rho_H, \sigma_A, \sigma_B)$
as before.  State point C' at $\rho_L$ and $\sigma_{C'}=\sigma_B$ has, however, a much smaller number of ions $N_{C'}=2\rho_L
V+A\Gamma(\rho_L,\sigma_B)$ than $N_B=N_C$ in state B and C, because its surface charge $\sigma_{C'}<\sigma_C$. Trajectory C'D' at fixed $\rho_L$ is
quite similar to CD but extends much further down to $\sigma_{D'}=\sigma_A$, where the number of ions in D' is even further reduced to the minimum
value in the cycle $N_{D'}/A=1.0$ nm$^{-2}$. Finally, at fixed $\sigma_{D'}$ the number of ions increases up to $N_A$ by gradually increasing $\rho_s$ from
$\rho_L$ to $\rho_H$. So also this cycle is completely determined by $\rho_H$, $\rho_L$, $\sigma_A$, and $\sigma_B$. The electric work $W'$ done
during the cycle ABC'D'A follows from Eq.(2) and reads $W'/A= 3.2 k_BT$ nm$^{-2}$, which is equivalent to $W'/\Delta N'=1.8 k_BT$ where $\Delta
N'=N_{B'}-N_{D'}$ is the number of ions that was exchanged between the two reservoirs during the cycle.

Clearly, $W'>W$, i.e. the Brogioli-type cycle with the ``isocharges'' BC' and D'A produces more work than the presently proposed ABCDA cycle with
canonical trajectories BC and DA. However, the efficiency of ABCDA, defined as $W/\Delta N$, indeed exceeds the efficiency $W'/\Delta N'$ of the
ABC'D'A cycle. This is also illustrated in Fig.\ref{fig:cyclus2}, where the two cycles ABCDA (a) and ABC'D'A (b) are shown in the $N$-$\mu$
representation. Whereas the total area of (b) is larger than that of (a), so $W'>W$ according to Eq.(\ref{W}), the larger spread in $\Delta N'$
compared to $\Delta N$ renders the efficiency of (b) smaller. The work $W'$ is therefore less than the decrease of the free energy of the reservoirs combined. The hatched area of Fig.\ref{fig:cyclus2}(b) denotes the work that could have been done with the number $\Delta N'$ of exchanged ions, if a cycle of the type ABCDA had been used.

The fact that $W'>W$ while $W/\Delta N > W'/\Delta N'$  proves to be the case for all charge densities $\sigma_A$ and $\sigma_B$ for which we
calculated $W$ (of an ABCDA-type cycle) and $W'$ (of an ABC'D'A-type cycle), at the same reservoirs $\rho_H$ and $\rho_L$ and the same pore size
$L$ as above. This is illustrated in Table 1, which lists $W$ and $W'$ per unit area and per transported ion for several choices of $\sigma_A$
and $\sigma_B$. The data of Table 1 shows that $W'>W$ by up to a factor 2, while $W/\Delta N > W'/\Delta N'$ by up to a factor of three for $V_o=0$, and a factor 8 for $V_o=V_e$. We thus conclude that the choice for a particular cycle to generate electric work depends on optimization considerations; our results show that maximum work or maximum efficiency do not necessarily coincide.

Table 1 not only shows the work per area and per ion, but in the last column also $W'/A\Delta\sigma$ with $\Delta \sigma=\sigma_B-\sigma_A$, {\em i.e.} the work per charge that is put on the electrode during the charging of trajectory AB. Interestingly, in these units the work is comparable to $\Delta\mu=3.3 k_BT$ provided $\sigma_A\gg\sigma^*$, as also follows from Gouy-Chapman theory for highly charged surfaces. Note that the work per transported charge does \emph{not} equal the amount of performed work per transported ion as $\Delta N / A$ is typically much larger than $\Delta \sigma$. Nevertheless, the fact that $W' \simeq \Delta \mu A (\sigma_B - \sigma_A)$ gives us a handle to link our results with the experiments of Brogioli \cite{brogioli}. During the experiment, the charge on the electrodes varies by $\delta Q = A e (\sigma_B - \sigma_A) \approx 0.25$ $m$C, such that one arrives at an expected work of 6 $\mu$J per electrode. This agrees reasonably well with the obtained value of 5 $\mu J$ out of the entire system. Unfortunately, the relation between the electrostatic potential and the charge in the experiments differs significantly from that of our theory by at least hundreds of millivolts; at comparable electrostatic potentials the charge density in Brogioli's experiments is almost two orders of magnitude smaller than our theoretical estimates. Therefore a qualitative comparison with the Brogioli-cycle is at this point very hard. The relatively low experimental charge densities clarify the lower amount of work produced per gram of electrode, which was noted earlier in the text. Including the Stern layer may be a key ingredient that is missing in the present analysis \cite{zhao}.

\begin{widetext}
\vspace*{0 cm}
\begin{table}[]
\begin{tabular}{ |c | c || c | c || c | c | c |}
\hline
$\sigma_A$ ($\rm{nm}^{-2}$)& $\sigma_B$ ($\rm{nm}^{-2}$)& $W/A$ ($k_BT$ / $\rm{nm}^2$)& $W/\Delta N$ ($k_BT$)& $W'/A$ ($k_BT$ / $\rm{nm}^2$)& $W'/\Delta N'$ ($k_BT$) & $W' / (A \Delta \sigma)$ ($k_BT$)\\
\hline
$1.0$ & $2.0$ & $2.3$ & $3.3$ & $3.2$ & $1.8$ -- $1.0$ & $3.2$\\
$1.0$ & $2.75$ & $4.9$ & $3.3$ & $5.6$ & $2.3$ -- $1.5$ & $3.2$\\
$0.5$ & $1.0$ & $1.1$ & $3.3$ & $1.5$ & $1.0$ -- $0.5$ & $3.0$\\
$0.1$ & $0.55$ & $0.6$ & $3.3$ & $1.1$ & $0.7$ -- $0.4$ & $2.4$\\
\hline
\end{tabular}
\caption{The work $W$ and $W'$ of cycles ABCDA and ABC'D'A, respectively, as illustrated in Figs.\ref{fig:cyclus}(a) and \ref{fig:cyclus2}, for
several choices of surface charges $\sigma_A$ and $\sigma_B$ in states A and B, for systems operating between electrolytes with high hand low salt
concentrations $\rho_H=0.6$ M and $\rho_L=0.024$ M, for electrodes with pore size $L=4$ nm. We converted $W$ and $W'$ to room temperature thermal
energy units $k_BT$, and not only express them per unit electrode area $A$ but also per exchanged number of ions $\Delta N$ and $\Delta N'$ during
the two cycles, respectively. Note that $W/\Delta N$ is a property of the two reservoirs, not of the charge densities of the cycle. Also note that $W'/\Delta N'$ depends on the volume $V_o$ of electrolyte outside the electrodes, here we successively give values for the optimal situation $V_o=0$ as well as for the situation $V_o = V_e$.}
\end{table}
\end{widetext}

\section{Limited fresh water supply}
Of course many more cycles are possible.  The two cycles ABCDA and ABC'D'A considered so far generate electric work out of the mixing of two very large reservoirs of salt and fresh water, taking up ions from high-salt water and releasing them in fresh water. Due to the large volume of the two reservoirs the ionic chemical potentials $\mu_H$ and $\mu_L$, and hence the bulk salt concentrations $\rho_H$ and $\rho_L$ in the reservoirs, do not change during this transfer of a finite number of ions during a cycle. However, there could be relevant cases where the power output of an osmo-electric device is limited by the finite inflow of fresh water, which then becomes brackish due to the mixing process; usually there is enough sea water to ignore the opposite effect that the sea would become less salty because of ion drainage by a cycle. In other words, the volume of fresh water cannot always be regarded as infinitely large while the salt water reservoir is still a genuine and infinitely large ion bath. The cycle with a limited fresh water supply is equivalent to a heat-engine that causes the temperature of its cold ``bath'' to rise due to the release of rest heat from a cycle, while the hot heat bath does not cool down due to its large volume or heat capacity. Here we describe and quantify a cycle ABCA that produces electric work by reversibly mixing a finite volume of fresh water with a reservoir of salt water.

\begin{figure}[]
\centering
\includegraphics[width = 7cm]{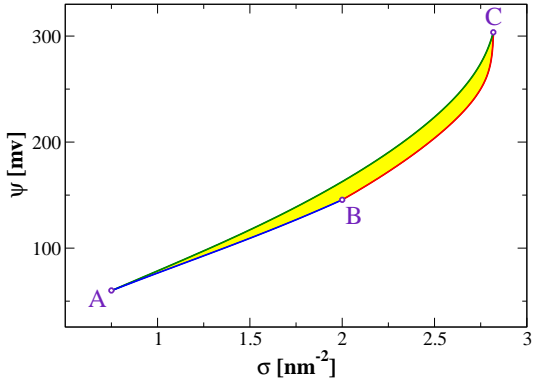}
\caption{Electrode potential $\Psi$ versus electrode charge density $\sigma$ representation of a charging-discharging cycle ABCA of an electrode with pore size $L=4$ nm (see text). The trajectory AB represents a grand-canonical charging process during which the system is connected to a ``sea'' with a fixed high salt concentration $\rho_H=0.6$ M, the electrolyte in the pores taking up ions. The number of ions during trajectory BC is kept fixed by disconnecting the electrodes from any ionic bath, such that further charging leads to more ionic adsorption at the expense of the bulk concentration, which has reduced to the fresh water concentration $\rho_L=0.024\rm{M}$ in C. Trajectory CA describes the discharging of the electrodes in contact with a finite volume $\Delta V=0.75 V$ of initially fresh water at concentration $\rho_L$, which becomes saltier due to the uptake of electrode-released ions until the sea concentration of salt $\rho_H$ is reached in A.} \label{fig:newcycle}
\end{figure}

We consider a finite volume $\Delta V=0.75V$ of fresh water with a low salt concentration $\rho_L=0.024$ M, such that the number of ions in this compartment equals $2\rho_L \Delta V$. This fresh water is assumed to be available at the beginning of a (new) cycle; its fate at the end of the cycle is to be as salty as the sea by having taken up $2(\rho_H-\rho_L)\Delta V$ ions from the electrode (which received them from the sea), with $\rho_H=0.6$ M the salt concentration in the sea. The cycle, which is represented in Fig.~\ref{fig:newcycle},  starts with the electrodes connected to a large volume of sea water at concentration $\rho_H$, charged up in state A at a charge density $\sigma_A=0.75$ nm$^{-2}$. During the first part AB of the cycle, the electrodes are further charged up until the positive one has taken up $2(\rho_H-\rho_L)\Delta V$ ions in its pores, which fixes the surface charge $\sigma_B=2.0$ nm$^{-2}$ in state B. Then the electrodes are to be disconnected from the sea, after which the charging proceeds in trajectory BC such that the increasing ion adsorption at a fixed total ion number reduces the salt chemical potential down to $\mu_L$ (and hence the salt concentration far from the electrode surface down to $\rho_L$) at $\sigma_C=2.8 $ nm$^{-2}$ in state C. The system can then be reversibly coupled to the finite compartment of initially fresh water, after which the discharging process CA takes place such that the released ions cause the fresh water to become more salty, reaching a charge density $\sigma_A$ when the salt concentration in the compartment of volume $\Delta V$ equals $\rho_H$. The cycle can then be repeated by replacing the compartment $\Delta V$ by fresh water again.

The relation between the surface potential $\Psi$, the charge density on the electrodes $\sigma$, and the ion reservoir concentration $\rho_s$ or the ion number $N$,  was numerically calculated using the modified PB-equation (\ref{modpb}) with BC's~(\ref{bc1}) and~(\ref{bc2}), combined with the adsorption relation (\ref{gam}), with the same parameters $v=a^3$, $a=0.55$ nm, $L=4$ nm, and $V=V_e=AL/2$ as before.   The enclosed area in Fig.~\ref{fig:newcycle} gives, using Eq.(\ref{W}), the net amount of (reversible) work $W$ performed during a cycle, which again equals the decrease of the free energy of the salt water reservoir and fresh water volume combined. In fact, this work can be calculated analytically as
\begin{equation}
W = \Delta V   \left[(\rho_H-\rho_L)k_BT  - \rho_L (\mu_H-\mu_L)\right], \label{newwork}
\end{equation}
where $\mu_H-\mu_L=k_BT\ln(\rho_H/\rho_L) $. This result agrees with the prediction by Pattle~\cite{pattle} for very small $\rho_L$.  For the parameters of the cycle discussed here, we find $W/\Delta V=1.2$ kJ per liter of fresh water, or $W/A= 0.45 k_BT$ nm$^{-2}$. The figures show that the amount of work per ion that is transported is typically smaller than what we
found for the Carnot-like cycle, of course.\\

We may compare this reversible cycle with the one proposed by Biesheuvel for the reverse process, which is called desalination. This cycle is very similar to ours, except that Biesheuvel's switching step from sea-to river water and v.v. is actually an iso-$\Gamma$ trajectory instead of our iso-$N$ tracjectory. This iso-adsorption trajectory does not seem to have a reversible heat engine analogue, as the degree of reversibility depends on the extent to which the electrolyte can be drained out of the micropores before. Nevertheless, we find agreement with the work that must be provided in the case of only a relatively small output volume of fresh water, and the expression found by Biesheuvel exactly equals Eq.~(\ref{newwork}). The point we would like to stress is that irreversible mixing during the switching step can be prevented by introducing a canonical(iso-$N$) part into the cycle which enables the system to adapt to a new salt concentration in a time-reversible fashion, such that maximal efficiency is preserved.

\section{Summary, conclusion and discussion}
Although substantial attempts to extract renewable energy from salinity gradients go back to the 1970's, there is considerable recent progress in this field stemming from the availability of high-quality membranes \cite{dlugolecki} and large-area nanoporous electrodes \cite{chmiola} with which economically interesting yields of the order of 1 kJ per liter of fresh water can be obtained ---equivalent to a waterfall of one hundred meter. The key concept in the recent experiments of Brogioli \cite{brogioli} is to cyclically (dis)charge a supercapacitor composed of two porous electrodes immersed in sea (river) water. In this article we have used a relatively simple density functional, based on mean-field electrostatics and a lattice-gas type description of ionic steric repulsions, to study the relation between the electrode potential $\Psi$, the electrode surface charge density $\sigma$, the ion adsorption $\Gamma$, the ion chemical potential $\mu$, and the total number of ions $N$ in a (slit-like) pore of width $L$ that should mimic the finite pores of the electrodes. With this microscopic information at hand, we have analysed several cycles of charging and discharging electrodes in sea and river water. By making an analogy with heat engines, for which the most-efficient cycle between two heat baths at fixed temperatures is the Carnot cycle with isothermal and  adiabatic (de)compressions, we considered cycles composed of iso$-\mu$ and iso$-N$ (dis)charging processes of the electrodes. We indeed found that these cycles are maximally efficient in the sense that the work per `consumed' ion that is transported from the sea to the river water during this cycle is optimal, given the salt concentrations in the river- and sea water. However, although the cycles used by Brogioli, with two iso$-\mu$ and two iso$-\sigma$ trajectories (where the latter are analogous to isochores in the heat-engine) are less efficient per transported ion, the total work of a ``Brogioli-cycle'' is larger, at least when comparing cycles that share the iso$-\mu$ charging in the sea water trajectory. We find, for electrode potentials $\Psi\simeq 100-300$ mV and electrode charge densities $\sigma\simeq 1-2$ nm$^{-3}$ in electrolytes with salt concentrations $\rho_H=0.6$ M (sea water) and $\rho_L=0.024$ M (river water), typical amounts of delivered work of the order of several $k_BT$ per transported ion, which is equivalent to several $k_BT$ per nm$^2$ of electrode area or several kJ per liter of consumed fresh water.\\

Our calculations on the Brogioli type of cycle agree with experiments regarding the amount of performed work per cycle with respect to the variance in the electrode charge during (dis-) charging; each unit charge is responsible for an amount of work that is given by the difference in chemical potential between the two reservoirs. However, the experimental data concerning the electrostatic potential could \emph{not} be mapped onto our numerical data. This could very well be due to the fact that the pore size in the experiments by Brogioli is very small such that ion desolvation, ion polarisability, and image charge effects may be determining the relation between the surface charge and electrostatic potential. Models which go beyond the present mean-field description are probably required for a quantitative description of this regime. Another ingredient in a more detailed description must involve the finite size of the ions combined with the microscopic roughness of the carbon. The ions in the solvent and the electrons (holes) in the electrode material cannot approach infinitely close, and the resulting charge free zone can be modeled by a Stern capacitance. Standard Gouy-Chapman-Stern (GCS) theory has successfully been applied to fit charge-voltage curves for porous carbon capacitive cells \cite{zhao, zhao2010} within the context of osmo-electrical and capacitive desalination devices. Extensions to GCS theory are currently being developed which include finite pore sizes, in order to obtain a physically realistic and simultaneously accurate model of the Stern layer within this geometry.
\\

Throughout this work we (implicitly) assumed the cycles to be reversible, which implies that the electrode (dis)charging is carried out sufficiently slowly for the ions to be in thermodynamic equilibrium with the instantaneous external potential imposed by the electrodes. This reversibility due to the slowness of the charging process has the advantage of giving rise to optimal conversion from ionic entropy to electric work in a given cycle. However, if one is interested in optimizing the {\em power} of a cycle, {\em i.e.} the performed work per unit time, then quasistatic processes are certainly not optimal because of their inherent slowness.  Heuristically one expects that the optimal power would result from the trade-off between reversibility (slowness) to optimize the work per cycle on the one hand, and fast electronic (dis)charging processes of the electrodes and fast fluid exchanges on time scales below the relaxation time of the ionic double layers on the other. An interesting issue is the diffusion of ions into (or out of) the porous electrode after switching on (or off) the electrode potential \cite{biesheuveldynamics, bazant}. Ongoing work in our group employs dynamic density functional theory \cite{marini1999, marini2000, archer2004} to find optimal-power conditions for the devices and cycles studied in this paper, e.g. focussing on the delay times between the electrode potential and the ionic charge cloud upon voltage ramps.\\

The recovery of useful energy from the otherwise definite entropy increase at estuaries, which may be relevant because our planet is so full of water, is just one example where one can directly build on Bob Evans' fundamental work on (dynamic) density functional theory, inhomogeneous liquids, electrolytes, interfaces, and adsorption.

\section{Acknowledgement}
It is a great pleasure to dedicate this paper to Bob Evans on the occasion of his 65th birthday. RvR had the privilege of being a postdoctoral fellow in Bob's group in the years 1997-1999 in Bristol, where he experienced an unsurpassed combination of warm hospitality, unlimited scientific freedom, and superb guidance on {\em any} aspect of scientific and (British) daily life. Bob's words ``Ren\'e, you look like a man who needs a beer!'' when entering the post-doc office in the late afternoon, which usually meant that he was ready to discuss physics after his long day of lecturing and administration, trigger memories of evening-long pub-discussions and actual pencil-and-paper calculations on hard-sphere demixing, like-charge attraction, liquid-crystal wetting, poles in the complex plane, or hydrophobic interactions, with (long) intermezzos of analyses of, say, Bergkamp's qualities versus those of Beckham. Even though not all of this ended up in publications, Bob's input, explanations, historic perspective, and style contained invaluable career-determining elements for a young postdoc working with him. RvR is very grateful for all this and more. We wish Bob, and of course also Margaret, many happy and healthy years to come.\\

We thank Marleen Kooiman and Maarten Biesheuvel for useful discussions. This work was financially supported by an NWO-ECHO grant.


\begin{thebibliography}{50}
\expandafter\ifx\csname natexlab\endcsname\relax\def\natexlab#1{#1}\fi
\expandafter\ifx\csname bibnamefont\endcsname\relax
  \def\bibnamefont#1{#1}\fi
\expandafter\ifx\csname bibfnamefont\endcsname\relax
  \def\bibfnamefont#1{#1}\fi
\expandafter\ifx\csname citenamefont\endcsname\relax
  \def\citenamefont#1{#1}\fi
\expandafter\ifx\csname url\endcsname\relax
  \def\url#1{\texttt{#1}}\fi
\expandafter\ifx\csname urlprefix\endcsname\relax\def\urlprefix{URL }\fi
\providecommand{\bibinfo}[2]{#2}
\providecommand{\eprint}[2][]{\url{#2}}

\bibitem[{\citenamefont{Pattle}(1954)}]{pattle}
\bibinfo{author}{\bibfnamefont{R.~E.} \bibnamefont{Pattle}},
  \bibinfo{journal}{Nature} \textbf{\bibinfo{volume}{174}},
  \bibinfo{pages}{660} (\bibinfo{year}{1954}).

\bibitem[{\citenamefont{Wick and Schmitt}(1977)}]{wick}
\bibinfo{author}{\bibfnamefont{G.~L.} \bibnamefont{Wick}} \bibnamefont{and}
  \bibinfo{author}{\bibfnamefont{W.~R.} \bibnamefont{Schmitt}},
  \bibinfo{journal}{Mar. Technol. Soc. J.} \textbf{\bibinfo{volume}{11}},
  \bibinfo{pages}{16} (\bibinfo{year}{1977}).

\bibitem[{\citenamefont{Levenspiel and de~Nevers}(1974)}]{levenspiel}
\bibinfo{author}{\bibfnamefont{O.}~\bibnamefont{Levenspiel}} \bibnamefont{and}
  \bibinfo{author}{\bibfnamefont{N.}~\bibnamefont{de~Nevers}},
  \bibinfo{journal}{Science} \textbf{\bibinfo{volume}{183}},
  \bibinfo{pages}{157} (\bibinfo{year}{1974}).

\bibitem[{\citenamefont{Loeb}(1976)}]{loeb1975}
\bibinfo{author}{\bibfnamefont{S.}~\bibnamefont{Loeb}}, \bibinfo{journal}{J.
  Membr. Sci.} \textbf{\bibinfo{volume}{1}}, \bibinfo{pages}{49}
  (\bibinfo{year}{1976}).

\bibitem[{\citenamefont{Loeb et~al.}(1976)\citenamefont{Loeb, Van~Hessen, and
  Shahaf}}]{loeb1976}
\bibinfo{author}{\bibfnamefont{S.}~\bibnamefont{Loeb}},
  \bibinfo{author}{\bibfnamefont{F.}~\bibnamefont{Van~Hessen}},
  \bibnamefont{and} \bibinfo{author}{\bibfnamefont{D.}~\bibnamefont{Shahaf}},
  \bibinfo{journal}{J. Membr. Sci.} \textbf{\bibinfo{volume}{1}},
  \bibinfo{pages}{249} (\bibinfo{year}{1976}).

\bibitem[{\citenamefont{Thorsen and Holt}(2009)}]{thorsen}
\bibinfo{author}{\bibfnamefont{T.}~\bibnamefont{Thorsen}} \bibnamefont{and}
  \bibinfo{author}{\bibfnamefont{T.}~\bibnamefont{Holt}}, \bibinfo{journal}{J.
  Membr. Sci.} \textbf{\bibinfo{volume}{335}}, \bibinfo{pages}{103}
  (\bibinfo{year}{2009}).

\bibitem[{\citenamefont{Post et~al.}(2007)\citenamefont{Post, Veerman,
  Hamelers, Euverink, Metz, Nymeijer, and Buisman}}]{post}
\bibinfo{author}{\bibfnamefont{J.~W.} \bibnamefont{Post}},
  \bibinfo{author}{\bibfnamefont{J.}~\bibnamefont{Veerman}},
  \bibinfo{author}{\bibfnamefont{H.~V.~M.} \bibnamefont{Hamelers}},
  \bibinfo{author}{\bibfnamefont{G.~J.~W.} \bibnamefont{Euverink}},
  \bibinfo{author}{\bibfnamefont{S.~J.} \bibnamefont{Metz}},
  \bibinfo{author}{\bibfnamefont{K.}~\bibnamefont{Nymeijer}}, \bibnamefont{and}
  \bibinfo{author}{\bibfnamefont{C.~J.~N.} \bibnamefont{Buisman}},
  \bibinfo{journal}{J. Membr. Sci.} \textbf{\bibinfo{volume}{288}},
  \bibinfo{pages}{218} (\bibinfo{year}{2007}).

\bibitem[{\citenamefont{Weinstein and Leitz}(1976)}]{weinstein}
\bibinfo{author}{\bibfnamefont{J.~N.} \bibnamefont{Weinstein}}
  \bibnamefont{and} \bibinfo{author}{\bibfnamefont{F.~B.} \bibnamefont{Leitz}},
  \bibinfo{journal}{Science} \textbf{\bibinfo{volume}{191}},
  \bibinfo{pages}{557} (\bibinfo{year}{1976}).

\bibitem[{\citenamefont{Veerman et~al.}(2009)\citenamefont{Veerman, Saakes,
  Metz, and Harmsen}}]{veerman}
\bibinfo{author}{\bibfnamefont{J.}~\bibnamefont{Veerman}},
  \bibinfo{author}{\bibfnamefont{M.}~\bibnamefont{Saakes}},
  \bibinfo{author}{\bibfnamefont{S.~J.} \bibnamefont{Metz}}, \bibnamefont{and}
  \bibinfo{author}{\bibfnamefont{G.~J.} \bibnamefont{Harmsen}},
  \bibinfo{journal}{J. Membr. Sci.} \textbf{\bibinfo{volume}{327}},
  \bibinfo{pages}{136} (\bibinfo{year}{2009}).

\bibitem[{\citenamefont{Post et~al.}(2008)\citenamefont{Post, Hamelers, and
  Buisman}}]{post2008}
\bibinfo{author}{\bibfnamefont{J.~W.} \bibnamefont{Post}},
  \bibinfo{author}{\bibfnamefont{H.~V.~M.} \bibnamefont{Hamelers}},
  \bibnamefont{and} \bibinfo{author}{\bibfnamefont{C.~J.~N.}
  \bibnamefont{Buisman}}, \bibinfo{journal}{Environ. Sci. Technol.}
  \textbf{\bibinfo{volume}{42}}, \bibinfo{pages}{5785} (\bibinfo{year}{2008}).

\bibitem[{\citenamefont{Dlugolecki et~al.}(2008)\citenamefont{Dlugolecki,
  Nymeijer, Metz, and Wessling}}]{dlugolecki}
\bibinfo{author}{\bibfnamefont{P.}~\bibnamefont{Dlugolecki}},
  \bibinfo{author}{\bibfnamefont{K.}~\bibnamefont{Nymeijer}},
  \bibinfo{author}{\bibfnamefont{S.~J.} \bibnamefont{Metz}}, \bibnamefont{and}
  \bibinfo{author}{\bibfnamefont{M.}~\bibnamefont{Wessling}},
  \bibinfo{journal}{J. Membr. Sci.} \textbf{\bibinfo{volume}{319}},
  \bibinfo{pages}{214} (\bibinfo{year}{2008}).

\bibitem[{sta()}]{statkraft}
\bibinfo{note}{See for example:
  \emph{http://www.statkraft.com/energy-sources/osmotic-power/}}.

\bibitem[{red()}]{redstack}
\bibinfo{note}{See for example:\\
  \emph{http://www.newscientist.com/article/dn19191-green-machine-a-salty-solu%
tion-for-power-generation.html}}.

\bibitem[{\citenamefont{Brogioli}(2009)}]{brogioli}
\bibinfo{author}{\bibfnamefont{D.}~\bibnamefont{Brogioli}},
  \bibinfo{journal}{Phys. Rev. Lett.} \textbf{\bibinfo{volume}{103}},
  \bibinfo{pages}{058501} (\bibinfo{year}{2009}).

\bibitem[{\citenamefont{Simon and Gogotsi}(2008)}]{simon}
\bibinfo{author}{\bibfnamefont{P.}~\bibnamefont{Simon}} \bibnamefont{and}
  \bibinfo{author}{\bibfnamefont{Y.}~\bibnamefont{Gogotsi}},
  \bibinfo{journal}{Nat. Mater.} \textbf{\bibinfo{volume}{7}},
  \bibinfo{pages}{845} (\bibinfo{year}{2008}).

\bibitem[{\citenamefont{Brogioli et~al.}(2011)\citenamefont{Brogioli, Zhao, and
  Biesheuvel}}]{zhao}
\bibinfo{author}{\bibfnamefont{D.}~\bibnamefont{Brogioli}},
  \bibinfo{author}{\bibfnamefont{R.}~\bibnamefont{Zhao}}, \bibnamefont{and}
  \bibinfo{author}{\bibfnamefont{P.~M.} \bibnamefont{Biesheuvel}},
  \bibinfo{journal}{Energy Environ. Sci.}  (\bibinfo{year}{2011}),
  \bibinfo{note}{accepted}.

\bibitem[{\citenamefont{Sales et~al.}(2010)\citenamefont{Sales, Saakes, Post,
  Buisman, Biesheuvel, and Hamelers}}]{sales}
\bibinfo{author}{\bibfnamefont{B.~B.} \bibnamefont{Sales}},
  \bibinfo{author}{\bibfnamefont{M.}~\bibnamefont{Saakes}},
  \bibinfo{author}{\bibfnamefont{J.~W.} \bibnamefont{Post}},
  \bibinfo{author}{\bibfnamefont{C.~J.~N.} \bibnamefont{Buisman}},
  \bibinfo{author}{\bibfnamefont{P.~M.} \bibnamefont{Biesheuvel}},
  \bibnamefont{and} \bibinfo{author}{\bibfnamefont{H.~V.~M.}
  \bibnamefont{Hamelers}}, \bibinfo{journal}{Environ. Sci. Technol.}
  \textbf{\bibinfo{volume}{44}}, \bibinfo{pages}{5661} (\bibinfo{year}{2010}).

\bibitem[{\citenamefont{Evans}(1979)}]{evans1979}
\bibinfo{author}{\bibfnamefont{R.}~\bibnamefont{Evans}}, \bibinfo{journal}{Adv.
  Phys.} \textbf{\bibinfo{volume}{28}}, \bibinfo{pages}{143}
  (\bibinfo{year}{1979}).

\bibitem[{\citenamefont{Tarazona et~al.}(1987)\citenamefont{Tarazona, Marini
  Bettolo~Marconi, and Evans}}]{tarazona1987}
\bibinfo{author}{\bibfnamefont{P.}~\bibnamefont{Tarazona}},
  \bibinfo{author}{\bibfnamefont{U.}~\bibnamefont{Marini Bettolo~Marconi}},
  \bibnamefont{and} \bibinfo{author}{\bibfnamefont{R.}~\bibnamefont{Evans}},
  \bibinfo{journal}{Mol. Phys.} \textbf{\bibinfo{volume}{60}},
  \bibinfo{pages}{573} (\bibinfo{year}{1987}).

\bibitem[{\citenamefont{Evans}(1989)}]{evans1989}
\bibinfo{author}{\bibfnamefont{R.}~\bibnamefont{Evans}},
  \emph{\bibinfo{title}{in ``Liquids and interfaces'', Les Houches Session
  XLVIII}} (\bibinfo{publisher}{Elsevier}, \bibinfo{year}{1989}).

\bibitem[{\citenamefont{Biesheuvel}(2009)}]{biesheuvel2009}
\bibinfo{author}{\bibfnamefont{P.~M.} \bibnamefont{Biesheuvel}},
  \bibinfo{journal}{J. Colloid Interface Sci.} \textbf{\bibinfo{volume}{332}},
  \bibinfo{pages}{258} (\bibinfo{year}{2009}).

\bibitem[{\citenamefont{Kittel}(1980)}]{kittel}
\bibinfo{author}{\bibfnamefont{C.}~\bibnamefont{Kittel}},
  \emph{\bibinfo{title}{Thermal Physics (2nd ed.)}} (\bibinfo{publisher}{W. H.
  Freeman and Company}, \bibinfo{address}{San Francisco and London},
  \bibinfo{year}{1980}).

\bibitem[{\citenamefont{Blundell and Blundell}(2006)}]{blundell}
\bibinfo{author}{\bibfnamefont{S.~J.} \bibnamefont{Blundell}} \bibnamefont{and}
  \bibinfo{author}{\bibfnamefont{K.~M.} \bibnamefont{Blundell}},
  \emph{\bibinfo{title}{Concepts in Thermal Physics}}
  (\bibinfo{publisher}{Oxford University Press}, \bibinfo{address}{Oxford},
  \bibinfo{year}{2006}).

\bibitem[{\citenamefont{Biesheuvel}()}]{biesheuvellaptop}
\bibinfo{author}{\bibfnamefont{P.~M.} \bibnamefont{Biesheuvel}},
  \emph{\bibinfo{title}{Private communication}}.

\bibitem[{\citenamefont{Tarazona}(1985)}]{tarazona1985}
\bibinfo{author}{\bibfnamefont{P.}~\bibnamefont{Tarazona}},
  \bibinfo{journal}{Phys. Rev. A} \textbf{\bibinfo{volume}{31}},
  \bibinfo{pages}{2672} (\bibinfo{year}{1985}).

\bibitem[{\citenamefont{Rosenfeld}(1989)}]{rosenfeld1989}
\bibinfo{author}{\bibfnamefont{Y.}~\bibnamefont{Rosenfeld}},
  \bibinfo{journal}{Phys. Rev. Lett.} \textbf{\bibinfo{volume}{63}},
  \bibinfo{pages}{980} (\bibinfo{year}{1989}).

\bibitem[{\citenamefont{Evans}(1991)}]{evans1991}
\bibinfo{author}{\bibfnamefont{R.}~\bibnamefont{Evans}},
  \emph{\bibinfo{title}{in ``Inhomogeneous Fluids''}}
  (\bibinfo{publisher}{Dekker}, \bibinfo{year}{1991}), \bibinfo{note}{edited by
  D. Henderson}.

\bibitem[{\citenamefont{Roth et~al.}(2002)\citenamefont{Roth, Evans, Lang, and
  Kahl}}]{roth2002}
\bibinfo{author}{\bibfnamefont{R.}~\bibnamefont{Roth}},
  \bibinfo{author}{\bibfnamefont{R.}~\bibnamefont{Evans}},
  \bibinfo{author}{\bibfnamefont{A.}~\bibnamefont{Lang}}, \bibnamefont{and}
  \bibinfo{author}{\bibfnamefont{G.}~\bibnamefont{Kahl}}, \bibinfo{journal}{J.
  Phys. Condens. Matter} \textbf{\bibinfo{volume}{14}}, \bibinfo{pages}{12063}
  (\bibinfo{year}{2002}).

\bibitem[{\citenamefont{Di~Caprio et~al.}(2007)\citenamefont{Di~Caprio,
  Valisk\'o, Holovko, and Boda}}]{dicaprio}
\bibinfo{author}{\bibfnamefont{D.}~\bibnamefont{Di~Caprio}},
  \bibinfo{author}{\bibfnamefont{M.}~\bibnamefont{Valisk\'o}},
  \bibinfo{author}{\bibfnamefont{M.}~\bibnamefont{Holovko}}, \bibnamefont{and}
  \bibinfo{author}{\bibfnamefont{D.}~\bibnamefont{Boda}}, \bibinfo{journal}{J.
  Phys. Chem. C} \textbf{\bibinfo{volume}{111}}, \bibinfo{pages}{15700}
  (\bibinfo{year}{2007}).

\bibitem[{\citenamefont{Bikerman}(1942)}]{bikerman}
\bibinfo{author}{\bibfnamefont{J.~J.} \bibnamefont{Bikerman}},
  \bibinfo{journal}{Philos. Mag.} \textbf{\bibinfo{volume}{33}},
  \bibinfo{pages}{384} (\bibinfo{year}{1942}).

\bibitem[{\citenamefont{Bazant et~al.}(2009)\citenamefont{Bazant, Kilic, D.,
  and Ajdari}}]{bazant2009}
\bibinfo{author}{\bibfnamefont{M.~Z.} \bibnamefont{Bazant}},
  \bibinfo{author}{\bibfnamefont{M.~S.} \bibnamefont{Kilic}},
  \bibinfo{author}{\bibfnamefont{S.~B.} \bibnamefont{D.}}, \bibnamefont{and}
  \bibinfo{author}{\bibfnamefont{A.}~\bibnamefont{Ajdari}},
  \bibinfo{journal}{Adv. Colloid Interface Sci.} pp. \bibinfo{pages}{48--88}
  (\bibinfo{year}{2009}).

\bibitem[{\citenamefont{Borukhov et~al.}(1997)\citenamefont{Borukhov, Andelman,
  and Orland}}]{borukhov}
\bibinfo{author}{\bibfnamefont{I.}~\bibnamefont{Borukhov}},
  \bibinfo{author}{\bibfnamefont{D.}~\bibnamefont{Andelman}}, \bibnamefont{and}
  \bibinfo{author}{\bibfnamefont{H.}~\bibnamefont{Orland}},
  \bibinfo{journal}{Phys. Rev. Lett.} \textbf{\bibinfo{volume}{79}},
  \bibinfo{pages}{435} (\bibinfo{year}{1997}).

\bibitem[{\citenamefont{Kralj-Iglic and Iglic}(1996)}]{kraljiglic}
\bibinfo{author}{\bibfnamefont{V.}~\bibnamefont{Kralj-Iglic}} \bibnamefont{and}
  \bibinfo{author}{\bibfnamefont{A.}~\bibnamefont{Iglic}}, \bibinfo{journal}{J.
  Phys. II France} \textbf{\bibinfo{volume}{6}}, \bibinfo{pages}{477}
  (\bibinfo{year}{1996}).

\bibitem[{\citenamefont{Paunov and Binks}(1999)}]{paunov}
\bibinfo{author}{\bibfnamefont{V.~N.} \bibnamefont{Paunov}} \bibnamefont{and}
  \bibinfo{author}{\bibfnamefont{B.~P.} \bibnamefont{Binks}},
  \bibinfo{journal}{Langmuir} \textbf{\bibinfo{volume}{15}},
  \bibinfo{pages}{2015} (\bibinfo{year}{1999}).

\bibitem[{\citenamefont{Paunov et~al.}(1966)\citenamefont{Paunov, Dimova,
  Kralchevsky, Broze, and Mehreteab}}]{paunov1996}
\bibinfo{author}{\bibfnamefont{V.~N.} \bibnamefont{Paunov}},
  \bibinfo{author}{\bibfnamefont{R.~I.} \bibnamefont{Dimova}},
  \bibinfo{author}{\bibfnamefont{P.~A.} \bibnamefont{Kralchevsky}},
  \bibinfo{author}{\bibfnamefont{G.}~\bibnamefont{Broze}}, \bibnamefont{and}
  \bibinfo{author}{\bibfnamefont{A.}~\bibnamefont{Mehreteab}},
  \bibinfo{journal}{J. Colloid Interface Sci.} \textbf{\bibinfo{volume}{182}},
  \bibinfo{pages}{239} (\bibinfo{year}{1966}).

\bibitem[{\citenamefont{Israelachvili}(1991)}]{israelachvili}
\bibinfo{author}{\bibfnamefont{J.}~\bibnamefont{Israelachvili}},
  \emph{\bibinfo{title}{Intermolecular and surface forces}}, vol.
  \bibinfo{volume}{2nd edition} (\bibinfo{publisher}{Academic Press},
  \bibinfo{address}{Amsterdam}, \bibinfo{year}{1991}).

\bibitem[{\citenamefont{van Roij}(2010)}]{vanroij2010}
\bibinfo{author}{\bibfnamefont{R.}~\bibnamefont{van Roij}},
  \bibinfo{journal}{Physica A} \textbf{\bibinfo{volume}{389}},
  \bibinfo{pages}{4317} (\bibinfo{year}{2010}).

\bibitem[{\citenamefont{Chapman.}(1913)}]{chapman}
\bibinfo{author}{\bibfnamefont{D.}~\bibnamefont{Chapman.}},
  \bibinfo{journal}{Philos. Mag.} \textbf{\bibinfo{volume}{25}},
  \bibinfo{pages}{475} (\bibinfo{year}{1913}).

\bibitem[{\citenamefont{Ettelaie and Buscall}(1995)}]{ettelaie}
\bibinfo{author}{\bibfnamefont{R.}~\bibnamefont{Ettelaie}} \bibnamefont{and}
  \bibinfo{author}{\bibfnamefont{R.}~\bibnamefont{Buscall}},
  \bibinfo{journal}{Adv. Colloid Interface Sci.} \textbf{\bibinfo{volume}{61}},
  \bibinfo{pages}{131} (\bibinfo{year}{1995}).

\bibitem[{\citenamefont{Biesheuvel}(2001)}]{biesheuvel2001}
\bibinfo{author}{\bibfnamefont{P.~M.} \bibnamefont{Biesheuvel}},
  \bibinfo{journal}{J. Colloid Interface Sci.} \textbf{\bibinfo{volume}{238}},
  \bibinfo{pages}{362} (\bibinfo{year}{2001}).

\bibitem[{\citenamefont{Verwey and Overbeek}(1948)}]{verweyoverbeek}
\bibinfo{author}{\bibfnamefont{E.~J.~W.} \bibnamefont{Verwey}}
  \bibnamefont{and} \bibinfo{author}{\bibfnamefont{J.~T.~G.}
  \bibnamefont{Overbeek}}, \emph{\bibinfo{title}{Theory of the Stability of
  Lyophobic Colloids}} (\bibinfo{publisher}{Elsevier}, \bibinfo{address}{New
  York}, \bibinfo{year}{1948}).

\bibitem[{\citenamefont{Ninham and Parsegian}(1971)}]{ninham}
\bibinfo{author}{\bibfnamefont{B.~W.} \bibnamefont{Ninham}} \bibnamefont{and}
  \bibinfo{author}{\bibfnamefont{V.~A.} \bibnamefont{Parsegian}},
  \bibinfo{journal}{J. Theor. Biol.} \textbf{\bibinfo{volume}{31}},
  \bibinfo{pages}{405} (\bibinfo{year}{1971}).

\bibitem[{\citenamefont{Torres et~al.}(2006)\citenamefont{Torres, van Roij, and
  Tellez}}]{torres}
\bibinfo{author}{\bibfnamefont{A.}~\bibnamefont{Torres}},
  \bibinfo{author}{\bibfnamefont{R.}~\bibnamefont{van Roij}}, \bibnamefont{and}
  \bibinfo{author}{\bibfnamefont{G.}~\bibnamefont{Tellez}},
  \bibinfo{journal}{J. Colloid Interface Sci.} \textbf{\bibinfo{volume}{301}},
  \bibinfo{pages}{176} (\bibinfo{year}{2006}).

\bibitem[{\citenamefont{Chmiola et~al.}(2006)\citenamefont{Chmiola, Yushin,
  Gogotsi, Portet, Simon, and Taberna}}]{chmiola}
\bibinfo{author}{\bibfnamefont{J.}~\bibnamefont{Chmiola}},
  \bibinfo{author}{\bibfnamefont{G.}~\bibnamefont{Yushin}},
  \bibinfo{author}{\bibfnamefont{Y.}~\bibnamefont{Gogotsi}},
  \bibinfo{author}{\bibfnamefont{C.}~\bibnamefont{Portet}},
  \bibinfo{author}{\bibfnamefont{P.}~\bibnamefont{Simon}}, \bibnamefont{and}
  \bibinfo{author}{\bibfnamefont{P.~L.} \bibnamefont{Taberna}},
  \bibinfo{journal}{Science} \textbf{\bibinfo{volume}{313}},
  \bibinfo{pages}{1760} (\bibinfo{year}{2006}).

\bibitem[{\citenamefont{Zhao et~al.}(2010)\citenamefont{Zhao, Biesheuvel,
  Miedema, Bruning, and van~der Wal}}]{zhao2010}
\bibinfo{author}{\bibfnamefont{R.}~\bibnamefont{Zhao}},
  \bibinfo{author}{\bibfnamefont{P.~M.} \bibnamefont{Biesheuvel}},
  \bibinfo{author}{\bibfnamefont{H.}~\bibnamefont{Miedema}},
  \bibinfo{author}{\bibfnamefont{H.}~\bibnamefont{Bruning}}, \bibnamefont{and}
  \bibinfo{author}{\bibfnamefont{A.}~\bibnamefont{van~der Wal}},
  \bibinfo{journal}{J. Phys. Chem. Lett.} \textbf{\bibinfo{volume}{1}},
  \bibinfo{pages}{205} (\bibinfo{year}{2010}).

\bibitem[{\citenamefont{Biesheuvel and Bazant}(2010)}]{biesheuveldynamics}
\bibinfo{author}{\bibfnamefont{P.~M.} \bibnamefont{Biesheuvel}}
  \bibnamefont{and} \bibinfo{author}{\bibfnamefont{M.~Z.}
  \bibnamefont{Bazant}}, \bibinfo{journal}{Phys. Rev. E}
  \textbf{\bibinfo{volume}{81}}, \bibinfo{pages}{031502}
  (\bibinfo{year}{2010}).

\bibitem[{\citenamefont{Bazant et~al.}(2004)\citenamefont{Bazant, Thornton, and
  Ajdari}}]{bazant}
\bibinfo{author}{\bibfnamefont{M.~Z.} \bibnamefont{Bazant}},
  \bibinfo{author}{\bibfnamefont{K.}~\bibnamefont{Thornton}}, \bibnamefont{and}
  \bibinfo{author}{\bibfnamefont{A.}~\bibnamefont{Ajdari}},
  \bibinfo{journal}{Phys. Rev. E} \textbf{\bibinfo{volume}{70}},
  \bibinfo{pages}{021506} (\bibinfo{year}{2004}).

\bibitem[{\citenamefont{Marini Bettolo~Marconi and
  Tarazona}(1999)}]{marini1999}
\bibinfo{author}{\bibfnamefont{U.}~\bibnamefont{Marini Bettolo~Marconi}}
  \bibnamefont{and} \bibinfo{author}{\bibfnamefont{P.}~\bibnamefont{Tarazona}},
  \bibinfo{journal}{J. Chem. Phys.} \textbf{\bibinfo{volume}{110}},
  \bibinfo{pages}{8032} (\bibinfo{year}{1999}).

\bibitem[{\citenamefont{Marini Bettolo~Marconi and
  Tarazona}(2000)}]{marini2000}
\bibinfo{author}{\bibfnamefont{U.}~\bibnamefont{Marini Bettolo~Marconi}}
  \bibnamefont{and} \bibinfo{author}{\bibfnamefont{P.}~\bibnamefont{Tarazona}},
  \bibinfo{journal}{J. Phys. Condens. Matter} \textbf{\bibinfo{volume}{12}},
  \bibinfo{pages}{A413} (\bibinfo{year}{2000}).

\bibitem[{\citenamefont{Archer and Evans}(2004)}]{archer2004}
\bibinfo{author}{\bibfnamefont{A.~J.} \bibnamefont{Archer}} \bibnamefont{and}
  \bibinfo{author}{\bibfnamefont{R.}~\bibnamefont{Evans}}, \bibinfo{journal}{J.
  Chem. Phys.} \textbf{\bibinfo{volume}{121}}, \bibinfo{pages}{4246}
  (\bibinfo{year}{2004}).

\end{thebibliography}
\end{document}